# Toward governance of cross-Cloud application deployment

Pierre de Leusse and Krzysztof Zieliński

Distributed System Research Group, AGH University of Science and Technology
Krakow, Poland
{pdl, kz}@agh.edu.pl

**Abstract.** In this article, the authors introduce the main ideas around the governance of cross-Cloud application deployment and their related concepts. It is argued that, due to the increasing complexity and nature of the Cloud market, an intermediary specialized in brokering the deployment of different components of a same application onto different Cloud products could both facilitate said deployment and in some cases improve its quality in terms of cost, security & reliability and QoS. In order to fulfill these objectives, the authors propose a high level architecture that relies on their previous work on governance of policy & rule driven distributed systems. This architecture aims at supplying five main functions of 1) translation of Service Level Agreements (SLAs) and pricing into a common shared DSL, 2) correlation of analytical data (e.g. monitoring, metering), 3) combination of Cloud products, 4) information from third parties regarding different aspects of Quality of Service (QoS) and 5) cross-Cloud application deployment specification and governance.

## 1  Introduction

Cloud computing has become an important part of software deployment and promises to offer virtually unlimited, cheaper, readily available, "utility type" computing resources. Many vendors have entered this market with different offerings ranging from Infrastructure as a Service (IaaS) such as Amazon Web Service (AWS) [1], to fully functional Platform as a Service (PaaS) such as the Google App Engine (GAE) [2] or Software as a Service (SaaS) like apigee [3]. In addition of different delivery models, Cloud products encompass today a very large set of services (e.g. storage, computation, security) that are provided using different technological environments and different characteristics such as pricing or deployment model (c.f. Figure 1 on too many choices for Cloud computing). As a result of this heterogeneity, deploying applications to a cloud and managing them often needs to be done using vendor specific methods. Investigation around breaking this "lock in" has seen a lot of activity over recent years both from academia and industry [4, 5] and practical solutions have started to appear [6, 7].

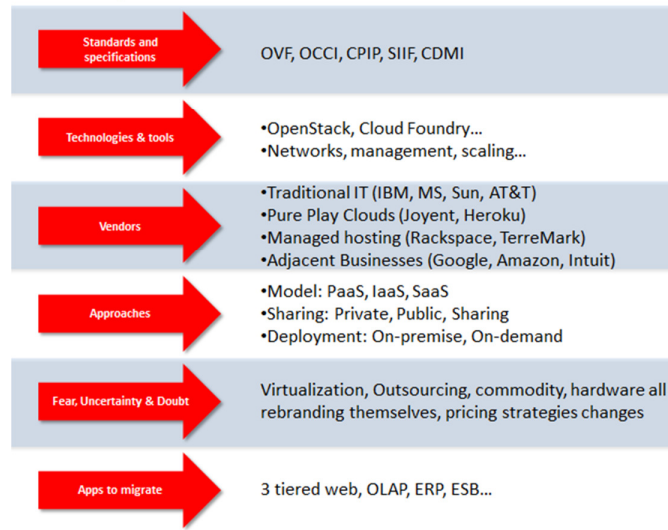

**Figure 1. Cloud computing: too many choices**

However, as Cloud usage and offerings continue to progress it has become both interesting and sometimes necessary to deploy different parts of a same software application over different Cloud products. This allows taking advantage of interesting opportunities such as different pricing strategies, security, reliability, elasticity, or performance provided by the different providers.

In this paper, the authors present the ideas, attached concepts and architecture of a governance infrastructure that efficiently automates efficient cross-Cloud application deployment strategies. This work leverages on the authors' previous works on SOA and policy & rule driven application governance [8, 9, 10, 11, 12] which use similar techniques.

In the section 2 "Pet clinic Grails application scenario", the authors describe a scenario that demonstrates the added value of cross-Clouds deployment governance. In section 3 "Requirements for a cross-Clouds application deployment governance broker" the authors define the challenges, concepts and designs considerations that such governance requires. In section 4 "Architecture of a cross-Clouds application deployment governance broker" a high level architecture of the Cloud Broker is described. In section 5 "Related work" the authors discuss the main recent advances in the domain of multiple Clouds management that enable the proposed approach. Finally, section 6 "Conclusions and future work" concludes and defines intended areas for further investigations.

It should be pointed out that this paper presents the concepts, design considerations and high level architecture. Implementation details of the proposed approach are considered beyond the scope of this paper and will be carried out along with experimentation and evaluation in future work.

## 2 Pet clinic Grails application scenario

In this section, the authors take a commonly used application, the Pet clinic application (PCA) and shortly describe how it could take advantage of a governed cross-Cloud deployment. The PCA is an online application that allows managing a pet clinic with the pets, their types and owner, the visits they make and the veterinaries they consult. Concretely, the PCA is a typical modern web application composed of several parts: database, domain model, business logic, presentation and logging. More details on this application and its content can be found at [13]. In comparison of the implementation aforementioned, we here take security into account and make use of the Spring security plugin for Grails [14]. This adds to the PCA with user (domain model) and authorization (business logic) related code. Figure 2 illustrates the anatomy of the PCA application in terms of Grails component.

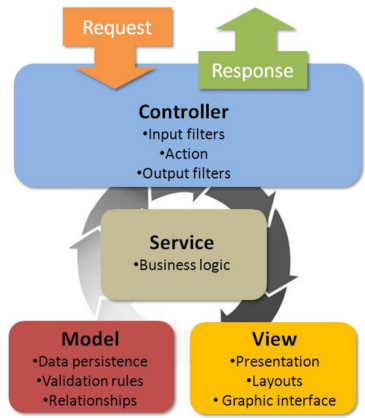

**Figure 2. Grails application anatomy**

In a traditional deployment, the PCA is deployed as a war file on a Java application server. More recently, a Grails plugin [15] has been released to allow for an easy integration with Cloud Foundry [16].

In this scenario, the PCA owner decided to see how practical and profitable it would be to run his application in a governed cross-Cloud manner. Using this approach, the PCA can take advantage of different pricing strategies in function of its needs and choose a more secured location for the database. As for most deployments, the PCA owner is required to specify a few Service Level Objectives (SLO) properties he wants to see enforced. In our context, he needs to specify, in a manifest, for the different parts of the application as he sees fit, characteristics regarding the price, the security & reliability and the performance.

Figure 3 depicts a sample part of the PCA Config.groovy file. This file is normally used by a Grails application to define different environment and configuration variables. On this figure, the different elements of the application are listed using a cross-Cloud deployment governance Domain Specific Language (DSL) together with the PCA's owner Cloud deployment choices. Concretely, the owner specifies from line 68 to 71 that the database named "prodDb", when the application is built for the

"production" environment, should be deployed on a private Cloud which URL, type and technology are provided. On lines 72 to 74, the owner chooses to deploy the domain classes in "economy" which, for the given DSL means the cheapest option available. On line 76, the three specified Controllers are required to be deployed according to the "bestEffort" option, which means a compromise between price and performance. The only Service of the PCA "SpringSecurityService" is required on lines 73-74 to be deployed using the same properties as the database.

With the SLOs defined the PCA owner can delegate his application deployment to a specialized entity: the cross-Cloud deployment governance broker, code named Cloud Broker (CB) henceforth.

Using the data presented on Figure 3, the CB is able to select the appropriate Cloud products and deploy the application. In order to achieve this, the CB is constantly gathering information about different Cloud offers available. These offers are stored, compared and sorting algorithms rank them using different characteristics such as location, pricing, performance, security and reliability.

In addition, using the "YieldGovernanceLifeCycleOptions.active" setting specified on line 67, the CB is taking an active role throughout the uptime of the PCA. For instance, when a Cloud provider SLA changes, it is compared again to the different "DeploymentOptions" and appropriate actions are taken. In this scenario, upon detecting that the provider used in the context of the "economy" option is no longer the best, the infrastructure in charge of governing the deployment could redeploy the relevant parts of the PCA and update the application in order to reflect these changes.

```
66  yieldCloudBroker = {
67      governance.lifecycle = YieldGovernanceLifeCycleOptions.active
68      dataSource = {
69          dataSources.environments[Environments.production].prodDb DeploymentOptions.privateCloud
70              ['http://149.156.97.139:9090', CloudTypes.paas, CloudProviders.OpenStackImagingService]
71      }
72      domainClasses = {
73          DomainClasses DeploymentOptions.economy
74      }
75      controllers = {
76          Controllers['Login', 'Logout', 'Pet'] DeploymentOptions.bestEffort
77          Controllers DeploymentOptions.economy
78      }
79      views = {
80          Views DeploymentOptions.economy
81      }
82      Services = {
83          Services[springSecurityService] DeploymentOptions.privateCloud
84              ['http://149.156.97.139:8080', CloudTypes.paas, CloudProviders.OpenStackImagingService]
85      }
86  }
```

**Figure 3. Config.groovy sample**

**Problem solved:** In the governed cross-Cloud deployment version of the PCA, the degree of involvement needed from the application owner is flexible. Indeed, the owner can specify the exact Cloud infrastructure and options that are required or simply identify the level of service needed and let the CB make the choice. With so many different providers, services and even types of Cloud, this can be a critical feature. This is pushed further as the CB constantly stays up to date with all the different offerings. As an added benefit, this means that the application owner can potentially save money on hosting, gain performance, reliability and/or security.

**Requirements:** In order for the CB to be possible and efficient, as many Cloud providers' offerings as possible should be discoverable and understandable and unambiguous (e.g. automated parsing of Service Level Agreement). In addition, seamless deployment on Cloud should be enabled for as many Cloud products as possible. Finally, in order to leverage as much as possible on such cross-Cloud deployment governance, applications need to be divided in self contained components as is possible for the Grails application. These issues are further discussed in the Related works section of this paper.

**Concerns:** The CB must manage the application based on the application owner's terms. It is therefore imperative that the cross-Cloud deployment governance DSL be understandable. Orthogonally, this DSL must be able to express the variations in SLA and QoS that are found on the market and that are defined using different units (e.g. CPU, memory, disk space, Dollars, Euros). In addition, for maximum efficiency, the CB must be able to pass contract with Cloud providers on behalf of the application owner. Cases where a list of Cloud providers along with secured login information can be given by the application owner could also made possible. Finally, the CB must be able to deploy an application without disclosing content or compromising the security of neither the application nor its owner.

In the next section, the requirements for each phase necessary to the governance of cross-Cloud application deployment are further discussed.

## 3 Requirements for a cross-Clouds application deployment governance broker

The objectives of the cross-Clouds application deployment governance broker code named Cloud Broker (CB) are to a) simplify the task of deploying software applications onto the Cloud and b) to increase the level of leverage that applications can expect from Cloud computing. Figure 4 illustrates the general anatomy of the CB, with the five main functions of 1) translation of Service Level Agreements (SLAs) and pricing, 2) correlation of analytical data (e.g. monitoring, metering), 3) combination of Cloud products, 4) information from third parties regarding different aspects of Quality of Service (QoS) and 5) application deployment governance.

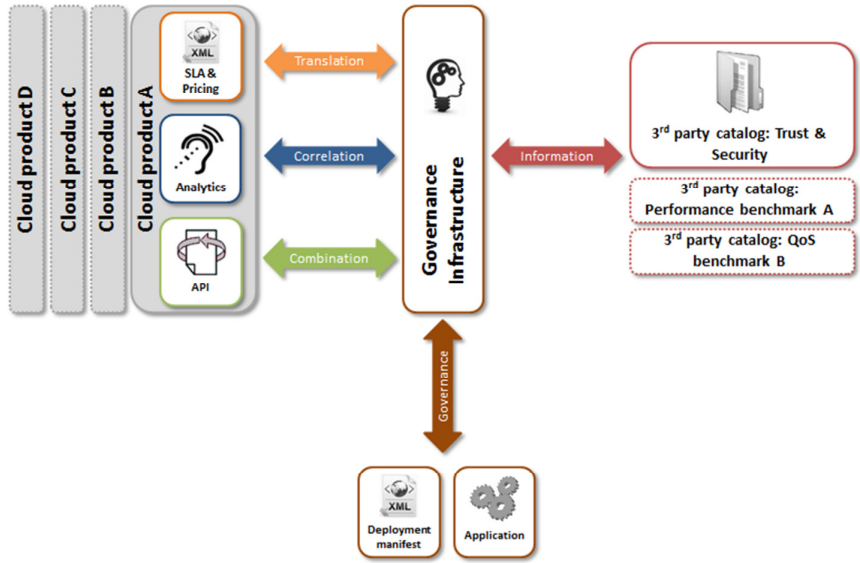

**Figure 4. Cross-Cloud application deployment governance infrastructure anatomy**

**Cloud product discovery**: In order to select Cloud products and deploy applications or their parts onto these products, it is necessary for the CB to know some of the Cloud characteristics. These features such as the type of service provided (e.g. storage, computation), the delivery method (e.g. SaaS, PaaS, IaaS), the interoperability, the portability, the API or the SLA need to be cataloged. More details on the important aspects that need to be known by the CB are described in [17, 21].
At the time of writing, the Cloud computing market is still very young and does not have any reliable directory. The discovery of Cloud products is left to a restricted group of initiates and suffers from a low transparency as the offers can strongly vary. However, as for most marketplaces, the authors anticipate that such registry service or survey will appear. A potential example of such catalog is Cloud Harmony [18] or such survey is the Gartner's Magic Quadrant for Cloud infrastructure as a service and Web hosting [19]. The subject of cataloging is discussed further in the latter point on Quality of Service information. In a first time, the CB will therefore make use of manually registered Cloud products with the possibility to automatically integrate with such public but potentially paying registries in the future. The Cloud products manually indexed in the CB registry will be registered in term of market volume and respect of standards and specifications.

**Translation of SLA and pricing:** Most Cloud providers communicate their pricing strategies and SLAs in a public manner. In order to propose such concepts as "Best effort" or "Economy" (c.f. Figure 3 and Section 2 Pet clinic Grails application scenario) the CB needs to compare Cloud products and offerings. The discovery aspects discussed in the previous point and the proposed catalog offer the basis for comparing Cloud products. However, the SLA and pricing may be based on different aspects of the Cloud (e.g. CPU, memory, network usage, database entry) for similar Cloud features and comparing both aspects together requires more treatments. It is noticeable that the authors do not take into account automatic SLA and pricing negotiation at this stage as this is not offered by any major Cloud provider.

**Correlation of analytical data:** it is impossible to know whether the terms of the SLA are being met without monitoring and measuring the performance of the service. Service Level Management is how that performance information is gathered and handled. Measurements of the service are based on the Service Level Objectives in the SLA or pricing strategies, depending on the Cloud provider, combined with the concrete Cloud resource usage and use of the application. Most Cloud products provide monitoring and/or measuring APIs (e.g. Amazon Web Service CloudWatch [20]), the goal of this correlation is to establish the relationship between the application usage, Cloud resource usage and the SLA and SLOs contracted in order to permit redeployment or use of the Cloud's elasticity when relevant (e.g. being overcharged, concrete QoS insufficient).

**Combination of Cloud:** the objective of the CB is to allow for an application Cloud deployment as seamless as possible. In order to do so, the CB must be able to use Cloud products regardless of their technologies and to potentially transfer applications from one onto another. The aspects of interoperability, portability and integration [21] are essential. Federation of Clouds is a very active research and development topic both in the industry and in academia. The most important works in this domain in the authors' view are discussed in the Related works section. It appears that, as the research and development of such technologies progress, together with the progress of standards and specifications for Cloud interoperability, it will become increasingly more manageable to deploy application regardless of the service provider. On the other hand, as new technologies continue to be developed it seems unrealistic to plan for a total interoperability between Cloud services. Finally, certain technological and architectural choices at the application level have an impact on what Cloud services are usable. For instance the Google App Engine (GAE) does not permit to deploy applications running on the .NET framework. Similarly, not every types of database can be deployed on any Cloud data store.

**Quality of Service information:** if the correlation of analytical data provides information for current deployments, it is also necessary to know in advance how different Cloud products can behave. For instance, a certain Cloud product will have a great QoS for users within its regional area but due to poor network connections will not be suitable otherwise. Some companies and organizations that provide testing services or test results are appearing [22] and can offer to the CB knowledge on Cloud products without having to use them. Other aspects than performance that are critical and need to be known before an application is deployed on a Cloud are security and reliability. Potential customers, including the CB can take better decisions on what Cloud provider to choose knowing how they handle on site security and backup. The choice of such third party information provider is critical and steps must be taken to insure that the data gathered in this manner is not bias. This capacity of information gathering is useful to the CB both in choosing Cloud providers but also in interpreting the analytics they provide.

**Application deployment governance:** finally, the CB's goal is to broker application deployment. In order to do so, application owner must provide their application together with a deployment manifest. This manifest, as illustrated on Figure 2, contains information about the deployment requirements for each part of the application. The level of details specified in the manifest can range from choosing a default option such as "Best Effort" to specifying the exact configuration parameters and location of the Cloud product required. The difficulty at the CB level is to provide a DSL that allows describing both end of this spectrum with equal ease of use for the application owner. In addition, this DSL must take into account all of the potential Cloud products and configurations possible. The added values of this approach are to 1) render the Cloud market more readable and make choices easier as well as 2) concretely prevent application owners from attempting to deploy an application on a inappropriate or wrongly configured Cloud product and thus increasing the speed and efficiency of the deployment. In addition, the manifest allows specifying the level of implication and adaption required from the CB throughout the application lifespan. In the proposed Pet clinic Grails application scenario, the application owner requires the CB to keep the deployment manifest choice true for the entire application life. This means that if a Cloud product disappears, changes its technology stack or pricing strategy the CB will attempt to adapt automatically.

## 4 Architecture of a cross-Clouds application deployment governance broker

The concepts and requirements defined in section 3 are taken into account in the proposed architecture as illustrated in Figure 5. The CB is divided into two main parts, the "Reference data" allows creating a global understanding of the Clouds products, the application deployment needs and the governance objectives. The "Governance logic" is the components and processes executed in order to enact and govern the deployment.

The metadata provided by the different external actors (e.g. Cloud providers, third party catalogs) provides information about the technical aspects of Cloud products (e.g. description of public APIs, type of product, method of delivery), pricing strategy (e.g. metering, plans) and QoS related data such as level of trust. Inside the CB, this is expressed using a common DSL. In addition, the client application that requires deployment governance provides a manifest that describes its needs (c.f. Figure 3). Finally, the governance is driven by its own policies that allow defining governance contexts. For instance, a governance policy could define that Cloud providers are ranked in the "economy" category when, in their own category (defined in the DSL), their prices on selected metrics are cheaper and their QoS in the context of a given application are still acceptable according to the deployment manifest.

The monitor point provides the mechanisms that collect, aggregate, filter, manage, and report monitoring details (metrics and topologies) collected from Cloud products. If a relevant change is detected it is passed to the decision point.

The decision point analyzes and correlates the reference data in order to take a deployment decision for a given deployment manifest.

The deployment plan is a collection that, for each application deployment, associates the application components with the selected Cloud products and the adequate configuration parameters.
The enforcement point provides the mechanisms that control the execution of a plan and deploy the components.

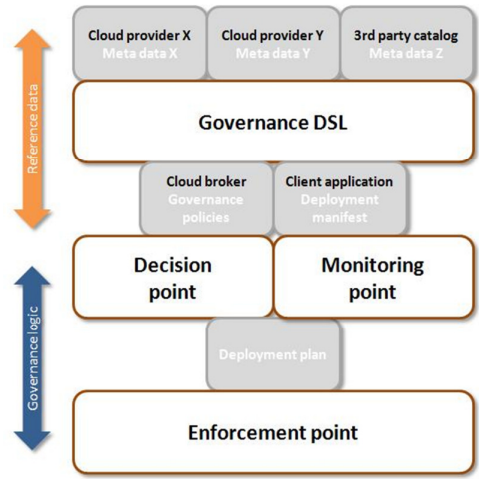

**Figure 5. Cross-Clouds application deployment governance broker high level architecture**

## 5 Related work

In this section we discuss other related approaches that aim at addressing some of the challenges of governing the deployment of application over multiple Clouds.
There are currently two main approaches to enabling multiple Clouds to work together, 1) providing Cloud products that are based an opened technologies and 2) providing management software that allows deploying Cloud components (e.g. application server, database, load balancer) onto multiple Clouds.
One way that can be seen as enabling both approaches is to leverage proposed standards and specifications such as the Open Virtualization Format (OVF) [25], the Open Cloud Computing Interface (OCCI) [26], the Cloud Portability and Interoperability Profiles (CPIP) [27], the Standard for Inter-Cloud Interoperability and Federation (SIIF) [28] or the Cloud Data Manage-ment Interface (CDMI) [29] to favor interoperability and to avoid being locked in to specific Cloud providers.
In the domain of opened technologies, sets of companies (e.g. IBM, Cisco, Dell, Intel, Microsoft) and communities are working together on OpenStack [5] in order to produce an open standard Cloud operating system for both public and private Clouds. At the time of writing, OpenStack consists of three core software projects: OpenStack Compute, OpenStack Object Storage and OpenStack Image.

In the domain of enabling multi-Clouds component deployment two approaches currently exist: Cloud brokerage and common Cloud middleware.

The European project Reservoir [23] is based on the concept of a Cloud federation, in which multiple IaaS Cloud providers (members of the federation) implement a common middleware layer that supports the operation of the federated Reservoir cloud. Because they share the same middleware technology, all providers of the Reservoir federation are able to communicate and cooperate (e.g. to migrate a virtual machine from one provider to another). At the time of writing, Reservoir does not support the integration of Cloud providers that implement a different middleware technology; neither offers any support to service modeling as a way to allow automatic creation of virtual appliances.

In a similar fashion, the IBM Altocumulus project [4] proposes a middleware platform that offers a unified API to deploy and manage application resources in heterogeneous clouds environments. Altocumulus services are mapped to the equivalent services offered by different cloud providers by means of Cloud-specific adapters. One major limitation of Altocumulus is that it defines its own unified API, which may discourage its adoption by Cloud users not interested in being locked to a proprietary technology. Another limitation of Altocumulus is that it does not yet support application deployment in a hybrid fashion. That is, with components belonging to the same virtual appliance being deployed in different clouds. In Altocumulus all virtual machines that are part of the same virtual appliance have to be deployed in the same IaaS Cloud.

Uni4Cloud [30] leverages on OVF and OCCI and aims at facilitating the deployment of Cloud components onto multiple Clouds using a model-based approach that helps to automatically configure and deploy applications independent of IaaS Cloud provider.

In the domain of multi-Clouds management and provisioning, RightScale [6], commercially provides a management platform for several commercial Cloud providers such as AWS, GoGrid and FlexiScale [31] and is currently developing support for different Cloud products.

Similarly, the open source project Aeolus [32] intends to provide a multi-Cloud management console and a unified API for managing packs of virtual machines across various private and public clouds.

The innovations introduced in this section go towards being able to deploy an application or Cloud components seamlessly onto different Cloud products. However, the features provided are always limited to the functional aspect of the deployment. In order to provide governance of cross-Clouds application deployment, it is necessary to be able to compare the different Cloud offers as introduced in section 3 on Anatomy of a cross-Cloud application deployment governance broker. To this day, this step is done manually by experts [22]. In [33] the authors specify that Cloud Service Brokerages (CSBs) are one of the most necessary and attainable opportunities for Cloud service providers. In this report, the authors explain how, due to the complex nature of the Cloud market, CSBs will broker relationships between a service consumer and a service provider in three different ways: 1) Cloud Service Intermediation by providing added services, 2) Cloud Service Aggregation by bringing together multiple services and 3) Cloud Service Arbitrage by providing seamless opportunistic choices.

## 6 Conclusions and future work

In this article, the authors introduce the main ideas around the governance of cross-Cloud application deployment and their related concepts. It is argued that, due to the increasing complexity and nature of the Cloud market, an intermediary specialized in brokering the deployment of different components of a same application onto different Cloud products could both facilitate said deployment and attempts to improve its quality in terms of cost, security & reliability and QoS. In order to fulfill these objectives, the authors propose a high level architecture that relies on their previous work on governance of policy & rule driven distributed systems. This architecture aims at supplying five main functions of 1) translation of Service Level Agreements (SLAs) and pricing into a common shared DSL, 2) correlation of analytical data (e.g. monitoring, metering), 3) combination of Cloud products, 4) information from third parties regarding different aspects of Quality of Service (QoS) and 5) application deployment governance.

As future work, a more complete analysis of our proposed architecture will be carried out. This will include an experimental evaluation of an implementation of the presented architecture. As a continuation to our work, it will be interesting to see how future work can not only facilitate the deployment of an application over different Cloud products but also improve and optimize the application and/or its components for particular deployments and Cloud products.

## References

1. Amazon Web Services (AWS), available at: http://aws.amazon.com/
2. Google App Engine (GAE), available at: http://code.google.com/appengine/
3. Apigee, embeddable APi console, available at: http://apigee.com/
4. Maximilien, E. M et al. 2009. IBM Altocumulus: A Cross-Cloud Middleware and Platform. In Proc. of the 24th ACM SIGPLAN Conference Companion on Object Oriented Programming Systems, Languages and Applications. OOPSLA'09. 805-806.
5. OpenStack, Open source software for building private and public clouds, available at: http://www.openstack.org/
6. Right Scale, Cloud management platform, available at: http://www.rightscale.com/
7. SMEStorage, Open Cloud Platform, available at: http://www.smestorage.com/
8. de Leusse, P., Kwolek, B. and Zielinski, K., A common interface for multi-rule-engine distributed systems, RuleML-2010 Challenge: 4th international rule challenge : Web rule symposium :Washington, DC, USA, October, 21–23, 2010, eds. Monica Palmirani, [et al.], http://ftp.informatic.rwth-aachen.de/Publications/CEUR-WS/
9. de Leusse, P., Kwolek, B. and Zielinski, K., A middleware infrastructure for multi-rule-engine distributed systems, Towards a service-based internet : third European conference, ServiceWave 2010 : Ghent, Belgium, December 13–15, 2010, eds. Elisabetta Di Nitto, Ramin Yahyapour. — Berlin ; Heidelberg : Springer-Verlag, cop. 2010. — (Lecture Notes in Computer Science ; ISSN 0302-9743 ; 6481), ISBN-10 3-642-17693-2, ISBN-13 978-3-642-17693-7.
10. de Leusse. P. and Dimitrakos, T., SOA-based security governance middleware, SECURWARE 2010: the fourth international conference on Emerging security information, systems and technologies : 18–25 July 2010, Venice, Italy, eds. Reijo Savola et al., IEEE, 2010, ISBN-13 978-0-7695-4095-5.
11. de Leusse, P., Dimitrakos, T., and Brossard, D., A governance model for SOA, IEEE 7th International Conference on Web Services (ICWS 2009), July 6-10, 2009, Los Angeles, CA, USA.
12. de Leusse, P. and Brossard, D., Distributed systems security governance, a SOA based approach, Third IFIP International Conference on Trust Management, Springer, June 15-19, 2009, Purdue University, West Lafayette, USA.
13. Rocher, G., The Grails PetClinic Application, A Grails Framework Demonstration, available at: http://petclinic-grails.cloudfoundry.com/html/petclinic.html
14. Beckwith, B., Spring Security Core Grails plugin, available at: http://www.grails.org/plugin/spring-security-core
15. Beckwith, B., Cloud Foundry Integration Grails plugin, available at: http://grails.org/plugin/cloud-foundry
16. Cloud Foundry, open platform as a service, available at: http://www.cloudfoundry.com/
17. Mell, P., Grance, T., The NIST Definition of Cloud Computing (Draft), Recommendations of the National Institute of Standards and Technology, NIST Special Publication 800-145 (Draft), January 2011
18. Cloud Harmony, benchmarking the Cloud, available at: http://cloudharmony.com/
19. Chamberlin, T., Leong, L., Gartner's Magic Quadrant for Cloud infrastructure as a service and Web hosting, Gartner RAS Core Research Note G00209074, 22 December 2010
20. Amazon CloudWatch, monitoring services for AWS cloud resources and the applications customers run on AWS, available at: http://aws.amazon.com/cloudwatch/
21. Cloud Computing Use Cases, A white paper produced by the Cloud Computing Use Case Discussion Group, Version 4.0, 2 July 2010
22. Comparethecloud, available at: http://www.comparethecloud.net/
23. Rochwerger, B. et al. 2009. The RESERVOIR Model And Architecture for Open Federated Cloud Computing. IBM Journal of Research and Development, 53(4):1-11, Special Edition on Internet Scale Data Centers.
24. GoGrid, Cloud Hosting and Hybrid Hosting offers cloud servers, dedicated servers, cloud storage, and F5 hardware load balancing, available at: http://www.gogrid.com/
25. Open Virtualization Format (OVF), available at: http://www.dmtf.org/standards/ovf
26. Open Cloud Computing Interface (OCCI), available at: http://occi-wg.org/
27. P2301 - Guide for Cloud Portability and Interoperability Profiles (CPIP), available at: http://standards.ieee.org/develop/project/2301.html
28. P2302 - Standard for Intercloud Interoperability and Federation (SIIF), available at: http://standards.ieee.org/develop/project/2302.html
29. Cloud Data Management Interface (CDMI), ), available at: http://cdmi.sniacloud.com/
30. Sampaio, A. and Mendonça, N, 2011, Uni4Cloud: an approach based on open standards for deployment and management of multi-cloud applications. In Proceeding of the 2nd international workshop on Software engineering for cloud computing (SECLOUD '11). ACM, New York, NY, USA, 15-21.
31. FlexiScale public Cloud, available at: http://www.flexiant.com/products/flexiscale/
32. Aeolus, Open source project for multi-Cloud management console and an unified API, available at: http://www.aeolusproject.org/
33. Plummer, D. C. and Kenney L. F., Three Types of Cloud Brokerages Will Enhance Cloud Services, Gartner RAS Core Research Note G00164265, 11 May 2009